# Clustered Volatility in Multiagent Dynamics

Michael Youssefmir and Bernardo A. Huberman

Dynamics of Computation Group
Xerox Palo Alto Research Center
Palo Alto, CA 94304

## Abstract

Large distributed multiagent systems are characterized by vast numbers of agents trying to gain access to limited resources in an unpredictable environment. Agents in these system continuously switch strategies in order to opportunistically find improvements in their utilities. We have analyzed the fluctuations around equilibrium that arise from strategy switching and discovered the existence of a new phenomenon. It consists of the appearance of sudden bursts of activity that punctuate the fixed point, and is due to an effective random walk consistent with overall stability. This clustered volatility is followed by relaxation to the fixed point but with different strategy mixes from the previous one. This phenomenon is quite general for systems in which agents explore strategies in search of local improvements.

## 1 Introduction

Large distributed multiagent systems are characterized by vast numbers of agents trying to gain access to limited resources in an unpredictable environment. This description applies to a wide range of systems, ranging from economies to computer networks and telecommunications. Because of this common characterization, insights and results from one type of multiagent system can at times be useful in another. For example, it has been pointed out several times that markets are not only of use in human societies, but can also be fruitful in the design and implementation of other large distributed systems[1–10].

In a distributed system, be it social or computational, agents do not always have complete information about the system in which they are embedded. This follows from their bounded rationality [11], i.e. the impossibility of choosing optimal strategies in a very complex setting. Bounded rationality stems from either lack of perfect information about the state of the system, or the inability on the part of agents to compute all possible outcomes in a timely fashion. Under such conditions, agents continuously switch between different behavioral modes in response to a constantly changing environment [12].

Studies of distributed systems have shown that when there is resource contention in the presence of delays, their dynamics is very complex, giving rise to nonlinear oscillations and chaos which drive the system far from optimality [13–16]. In some cases one can still observe a stable equilibrium . In the case of distributed computation, this undesirable behavior can be controlled through the introduction of local reward procedures, whereby the relative number of agents following effective strategies is increased at the expense of the others [17]. This procedure, which generates a diverse population out of an essentially homogeneous one, is able to control chaos in distributed computational systems through a series of dynamical bifurcations into a stable fixed point.

In this paper we study the effect of fluctuations on the model originally proposed by Hogg and Huberman [17] to analyze and control a distributed computational system. The model incorporates two essential elements that govern local decisions by the agents: uncertainty and strategies. Uncertainty, as discussed above, is related to bounded rationality. Strategies are behavioral modes on the part of agents, behaviors that are used to gain access to resources contested by other agents. Since agents can switch between strategies, the system can evolve into a multitude of different configurations, some of which can lead to enhanced adaptability as well as stability [18].

When we analyzed the fluctuations that arise from the fact that agents can temporarily change strategies, we discovered the existence of a new phenomenon. This phenomenon appears whenever the control mechanism is active and the system is close to equilibrium. As the dynamics of the multiagent system unfolds, the equilibrium state becomes punctuated by episodes of clustered volatility that take place at random. These bursts are caused by an effective random walk in the space of strategies, and they appear even in the case of large numbers of agents. This random walk is the resultant of a degeneracy in which many strategy mixes are consistent with overall stability. We also speculate that this new phenomenon may explain the clustered volatility observed in some economic time series, such as stock returns and exchange rates. In this context volatility clustering refers to the statistical evidence that periods of higher than normal fluctuations in the data tend to be clustered together [19, 20].

In the next section, we describe the model in detail and show how the introduction of diverse strategies leads to equilibrium. In Section 3 we perform computer simulations of a multiagent system allowing for strategy switching and show the appearance of clustered volatility. We also establish that this new phenomenon is not due to standard finite size effects, which decrease with the number of agents [21]. In section 4 we exhibit this random walk for the case of few strategies, and we determine the analytical criteria for its appearance. Finally, in section 5, we show the precise nature of the degeneracy in the model that leads to this random walk.

## 2 The Model

The model we consider attempts to capture the essential features of distributed systems consisting of intentional agents that adaptively react to the dynamics that unfold around them. Each agent acts in order to optimize its own utility. More specifically, we focus on the case in which agents have incomplete understanding and reasoning about the underlying model of the entire system. The interesting question is the extent to which imperfect knowledge on the part of agents can lead to coordination of the system as a whole.

In the model agents accrue utility by making choices between several resources. Each resource, *r*, has a performance metric or utility associated with it, which we denote by $G_r$. This utility depends on the fraction, $f_r$, of agents using that resource. If resource usage is competitive, then the associated payoff function will tend to decrease with the number of agents using that resource. On the other hand, some resources could have a cooperative function so the overall group performance is enhanced as more agents use that resource. In this case, the payoff function for this resource will tend to increase with the fraction of agents using it. For the sake of simplicity in this paper, we will limit the number of resources to two.

In order to make resource choices, agents need to estimate the relative payoffs of the various resources. Since agents do not have perfect information about the use and utility of a particular resource, we make a distinction between the actual payoffs, $G_r(f_r)$, that agents receive when they access them, and the perceived payoffs, denoted by $\hat{G}_r$, that they have before accessing them. This perceived payoff models the variability in an agent's ability to obtain exact information about the system. Uncertainty is taken into account by assuming that the perceived payoffs to be normally distributed around the actual payoff $G_r$, with a standard deviation, $\sigma$, i.e. $\hat{G}_r \sim N(G_r, \sigma)$. For simplicity, we will assume that all perceived payoffs have the same standard deviation.

From the dynamical point of view, there is a specific rate at which agents reevaluate their resource choices. We will assume that these reevaluations are asynchronous and statistical in nature, and determined by a Poisson rate $\alpha$. In an infinitesimal time increment $\Delta t$, the probability that a particular agent reevaluates its choice is then given by $\alpha \Delta t$. In the simplest case where strategies are not a factor in agent behavior, all agents process information about the resource utilities in the same way, i.e., they are given $\hat{G}_1$ and $\hat{G}_2$, and they pick the resource that maximizes utility.

**Dynamics without Strategies** We first consider the dynamics of resource utilization in the system in the limit of a large number of agents when agents do not use strategies. In this case, ignoring fluctuations that are suppressed inversely with the number of agents [21], the dynamics of the system for two resources reduces to the simple equation [13]

$$\frac{df}{dt} = -\alpha(f - \rho(f)) \tag{1}$$

where $f$ is the fraction of agents using the first resource and $\rho$ is the probability that an agent will prefer the first resource over the second when it makes a choice. In terms of the payoffs and uncertainty it is given by

$$\rho(f) = \frac{1}{2}\left(1 + \text{erf}\left(\frac{G_1(f) - G_2(1-f)}{2\sigma}\right)\right) \tag{2}$$

where, again, $\sigma$ quantifies the uncertainty in resource performance. Since the total number of agents is taken to be a constant,

$$f_2 = 1 - f_1. \tag{3}$$

Finally, the delays with which agents receive information about the resource usages enters by replacing the payoff arguments in eqn. 2 by their corresponding delayed values $f(t) \rightarrow f(t-\tau)$.

In the case of zero uncertainty and delay, i.e. $\sigma = 0$ and $\tau = 0$, agents are perfectly aware of the state of the system and are able to rapidly optimize their performance. In this case, resource contention in the system converges to a Nash equilibrium such that the resource payoffs are equal for all agents. As the uncertainty about payoffs increases, the equilibrium gradually moves away from the optimum state, approaching a situation in which equal fractions of agents use each resource. At the same time, the stochastic fluctuations associated with the probabilistic nature of perceived payoffs become more and more noticeable. Furthermore, if delays in information appear in the system, the solutions of eqn. 1 exhibit nonlinear and possibly chaotic oscillations, with a consequent degradation in the performance of the system.

**Dynamics with Strategies** The oscillatory and chaotic departures from the fixed point can be brought back into equilibrium by using a mechanism originally proposed by Hogg and Huberman [17]. It relies on allowing agents an extra dimension of behavior in order to choose among different strategies when evaluating information about the resources that they contend for. Each agent independently picks among the set of possible strategies according to how beneficial that strategy is perceived to be. Strategies thus encapsulate the various ways in which agents get and use information about the system in order to maximize their own performance. In this paper we consider two types of possible strategies.

The first type of strategy involves agents making decisions based on the past behavior of the system available to them. Such strategies may allow agents to take advantage of periodic behavior in the system, to act in a trend following manner, or to act as a contrarian moving against the trend. Such strategies might, for example, involve taking linear combinations of past values in order to obtain a prediction of future performance for the various resources. Specifically, considering the simplest such strategies, an agent making a decision based on the strategy labeled by the letter *s*, will use information in the system delayed by a time $\tau_s$. So, for example, an agent using strategy *s*, will choose to pick the first resource with probability $\rho_s = \rho(f(t - \tau_s))$.

The second set of strategies we consider in this paper are strategies that simply prefer one resource over another. In particular, an agent using strategy *s*, will choose to use the first resource with a probability

$$\rho_s = \frac{1}{2}\left(1 + \mathrm{erf}\left(\frac{G_1(f) - G_2(1-f) + b_s}{2\sigma}\right)\right) \tag{4}$$

where $b_s$ quantifies the strategy's preference to the first resource.

The state of the system is thus characterized not only by the fractions using each resource, $f_r^{res}$, but also by the fraction of agents using resource *r* and strategy *s*, $f_{rs}$. We will further define $f_s^{str}$ to be the fraction of agents using strategy *s* and call this the *strategy mix* in the system.

The Hogg-Huberman mechanism has agents continually searching for an optimal strategy as the dynamics of the system unfolds. In this way strategies that are obtaining the most useful information about the system increase in the population at the expense of others. For example, strategy optimization might involve agents observing the behavior and success of a few other agents and then picking a good strategy based on that information; or perhaps, agents keep records of how well strategies have performed in the past and make decisions based on that information. In either case, and indeed quite generally, the net effect of the strategy optimization procedure will be to change the usage of a certain strategy according to its overall performance. Note that we have thus explicitly assumed that agents do not have knowledge about which of the possible strategies is best. Instead, they make their strategy choices as the system continues to evolve. The strategy switching mechanism can thus be thought of as a way of learning in a distributed manner, with no single agent having full knowledge of the underlying model or state of the entire system.

In analogy with the resource reevaluation rate $\alpha$, we define another Poisson rate, $\gamma$, which agents make strategy choices. We will call $\gamma$ the *strategy switching rate* .. When an agent chooses a strategy *s* it does so with a probability denoted by $\eta_s$. $\eta_s$ satisfies the consistency requirement $\sum_s \eta_s = 1$, and is taken to be,

$$\eta_s = \frac{\sum_r f_{rs} G_r}{\sum_r f_r^{res} G_r} \tag{5}$$

The denominator in this expression is simply a measure of the total utility in the system, while the numerator is the amount of this total utility obtained by agents using strategy *s*. This form of $\eta_s$ guarantees the growth of strategies that create the most payoff in the system. Note that agents are assumed to make resource and strategy decisions separately and asynchronously. Moreover, we have assumed that agent strategy decisions are independent of which resource they are using. This assumption is very natural if agents switch resources regularly.

An example of the Hogg-Huberman mechanism in action is provided in figure 1 and figure 2, in which 99% of the agents are initially using the strategy with lowest delay ie. $\tau = 5$. This result, obtained by solving the dynamical equations, represents the average expected behavior in the system. As such, it ignores stochastic fluctuations that become less important as the number of agents in the system is increased. It shows the remarkable stabilization of an initially unstable system once strategy switching is allowed. As can be seen initially the system is quite chaotic; however, once agents are allowed to optimize and switch strategies over a large enough set of delays, the system finds an equilibrium distribution of strategies (depicted in figure 2) in which local agent behaviors at the microscopic level lead to a macroscopic equilibrium. The system is also robust in that sudden perturbations to the system eventually relax back to equilibrium [17].

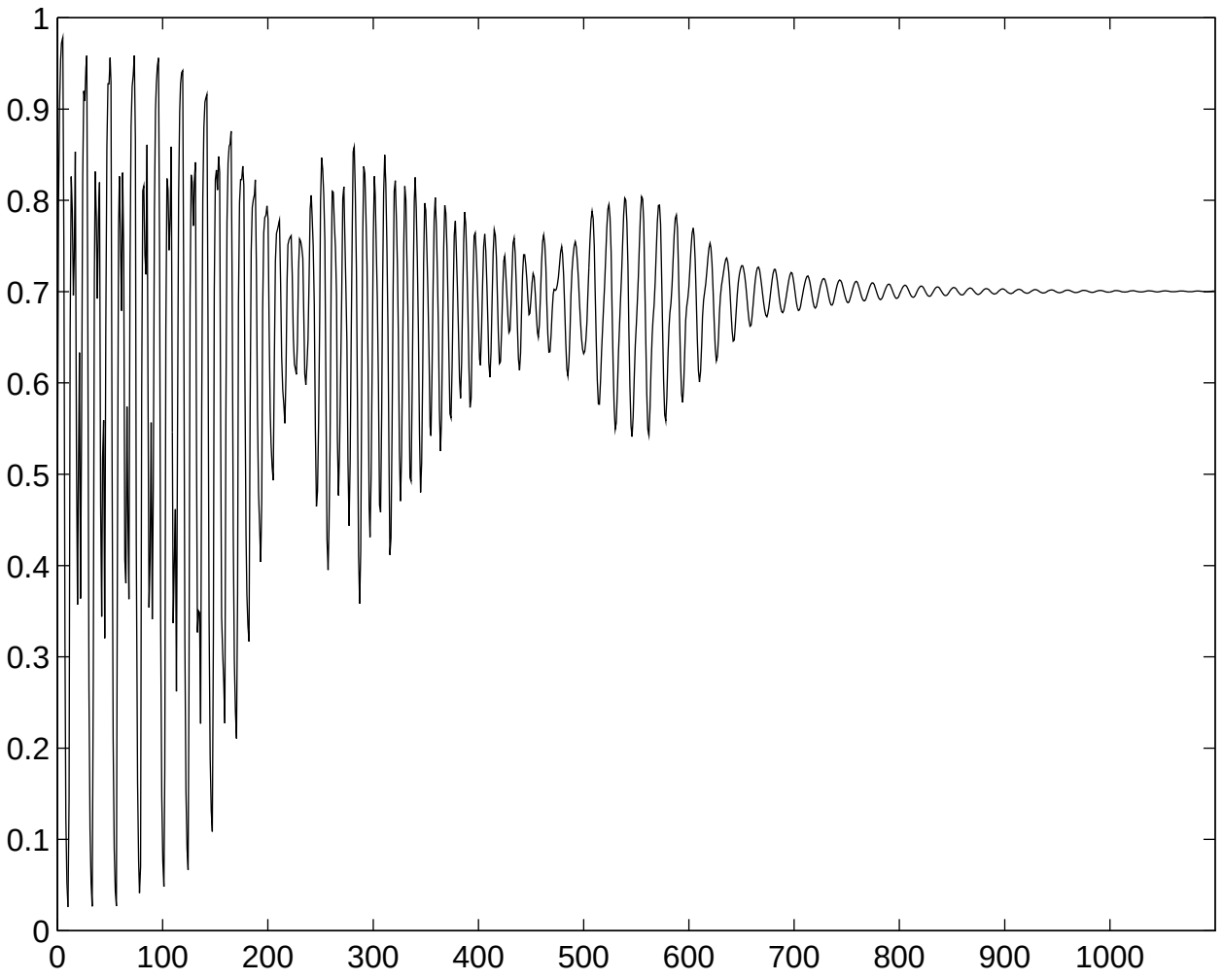

**Fig. 1.** Typical behavior of the Hogg—Huberman control mechanism [17] showing stabilization of resource contention when agents are allowed to switch strategies. Plot shows the fraction *f* of agents using resource 1 as a function of time, obtained by solving a set of equations (valid in the limit as the number of agents becomes large) [17]. Payoff functions are given by $G_1(f) = 4 + 7f - 5.333f^2$ and $G_2(f) = 7 - 3f$ together with parameters $\alpha = 1$ and $\sigma = 0.25$. Agents switch among 100 strategies at rate $\gamma = 1$ with the result that after transient oscillations the system settles into a state of stability. The 100 strategies are uniformly separated between $\tau = 5$ and $\tau = 45$ and initially 99% of agents are using the strategy with lowest delay.

When strategy choices are included, the simple eqn. 1 must be modified. The generalization of eqn. 1 is [17],

$$\frac{df_{rs}}{dt} = \alpha\big(f_s^{str}\rho_{rs} - f_{rs}\big) + \gamma(f_r^{res}\eta_s - f_{rs}) \qquad (6)$$

where $\rho_{rs}$ is the probability that an agent using strategy $s$ chooses to use resource $r$, and $\eta_s$ is the probability that an agent will switch to strategy *s*. The interpretation of eqns. 6 is facilitated by summing them over all the resources and then again over all strategies. The equations describing resource and strategy switching are then given by,

$$\frac{df_r^{res}}{dt} = -\alpha\left(f_r^{res} - \sum_s f_s^{str}\rho_{rs}\right) \tag{7}$$

$$\frac{df_s^{str}}{dt} = -\gamma\left(f_s^{str} - \eta_s\right) \tag{8}$$

The term preceded by $\alpha$ on the right hand side of eqn. 6 is analogous to the right hand side terms in eqn. 1, with $\rho_{rs}$ being the probability that an agent using strategy $s$ chooses to use resource $r$. Eqn. 8 represents the effects of strategy switching among the distribution of strategies within the agent population. This strategy mix affects the overall resource allocation in the system through eqn. 7.

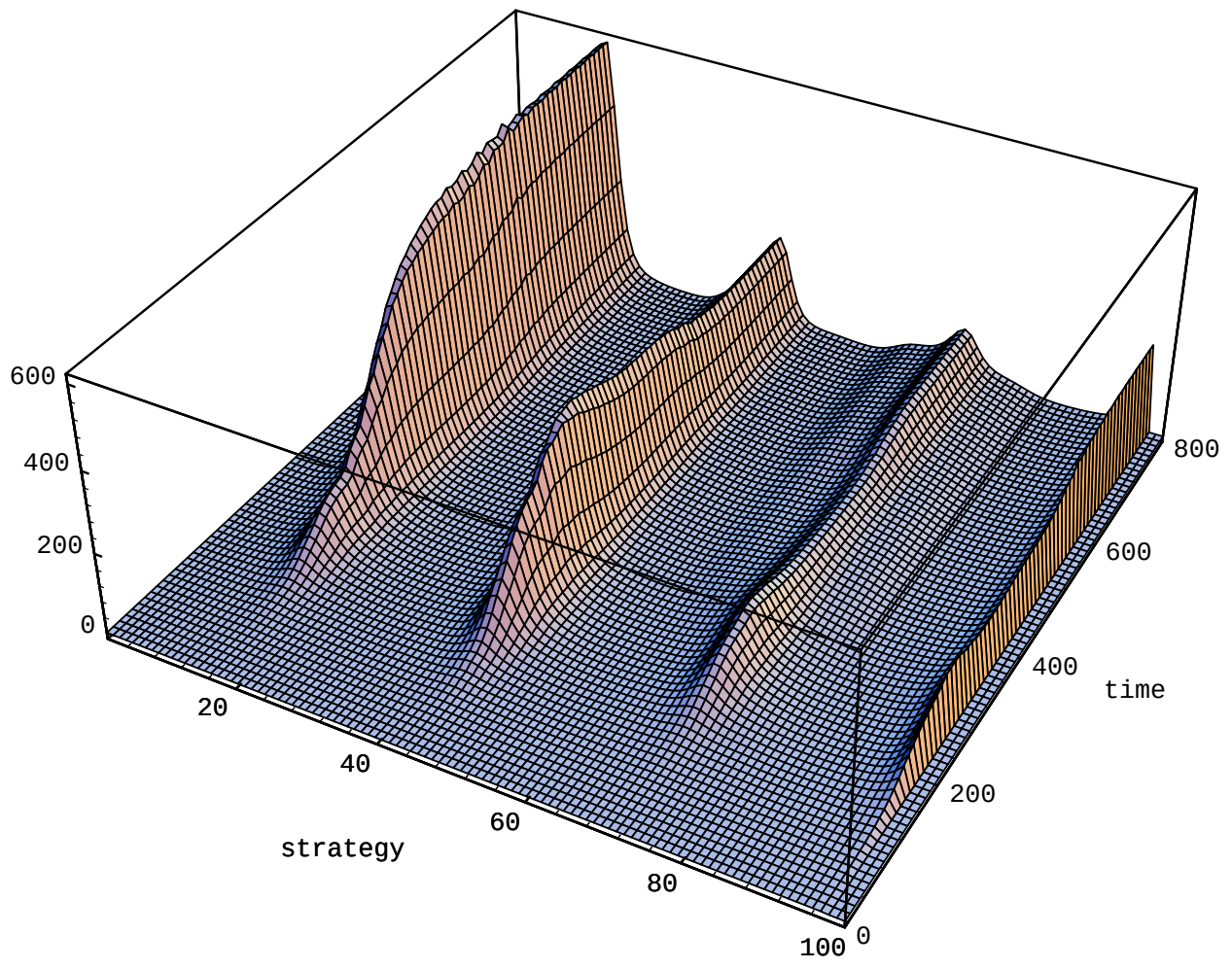

**Fig. 2.** The time evolution of the distribution of strategies for the simulation in figure 1. Vertical axis denotes the ratio $f_s^{str}(t)/f_s^{str}(0)$ where initially 99% are using the strategy with lowest delay. Strategies use delays that are uniformly spread between $\tau = 5$ and $\tau = 45$. For long times, the system evolves into a steady state distribution of strategies that is not unique.

## 3 Clustered Volatility

We should point that even when the resource utilizations come to a steady equilibrium, agents continue to vary their strategies in order to locally increase their utilities. In this way the system remains continually adaptive to the environment around it. For example, resource payoffs may suddenly change drastically, in which case the system may not be able to maintain stability if the strategy mix is frozen in time. Alternatively, resource payoffs might gradually change over time in which case the system must be continuously responsive to changes in the payoffs.

While the results we presented in figures 1 and 2 were obtained analytically, it is of interest to perform computer experiments to observe the stabilizing effect of strategy switching. The results of such an experiment are seen in figure 3 for a typical simulation of 5000 agents. The parameters, payoffs, and strategies are the same as in figure 1. The reward mechanism in the simulations has been slightly modified from that in eqn. 5 to insure that each strategy always maintains a minimal population of 5 agents. Comparison with the theoretical result presented in figure 1, shows that there are qualitatively different features that occur in the experiment. The simulations show the appearance of bursts of activity superimposed on a background of small fluctuations around the predicted equilibrium value. While the small background fluctuations are the result of the fact that agent decisions are themselves probabilistic, the larger bursts are the result of a more subtle mechanism, which we elucidate below.

To stress the generality of this type of behavior we show in figure 4 the results of another computer experiment in which all agents use information with a delay of $\tau = 10$ and in which strategies correspond to forming preferences towards the various resources. The simulation again consists of 5000 agents with 20 strategies. The corresponding preferences in eqn. 4, $b_s$, are uniformly chosen from the interval $[-1, 1]$. In the case of strategies based on the preferences towards particular resources, numerical integration of the eqns. $f_{rs}$ show that given this set of possible strategies, the system achieves stability at resource utilizations given by allocation among the resources given by $G_1(f_1) = G_2(f_2)$. Once again, the simulation shows that even when the number of agents is large the system exhibits bursts of activity.

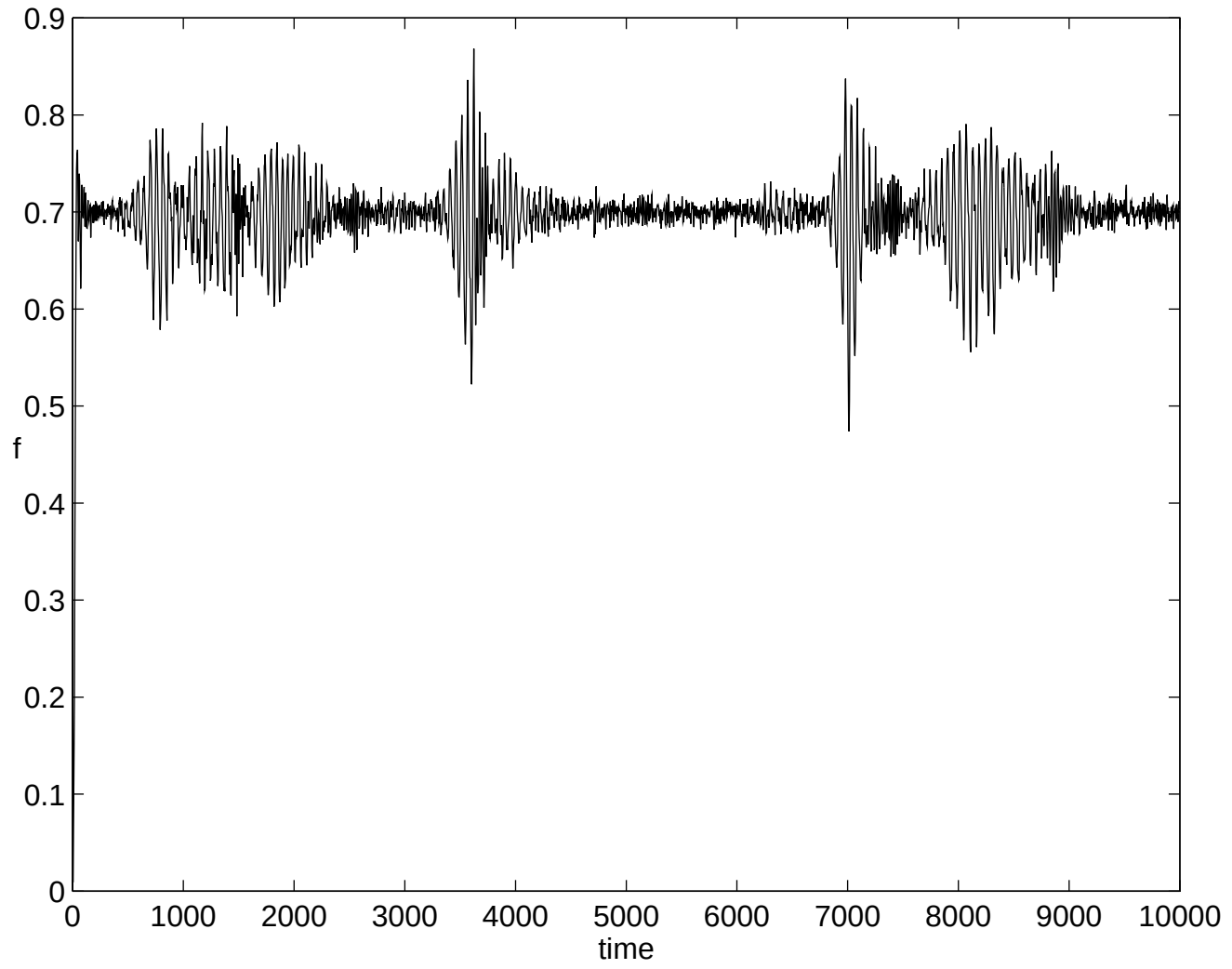

**Fig. 3.** Typical behavior seen in computer simulations of contention over two resources. The payoffs, parameters, and possible strategies are as in figure 1. The simulation consists of 5000 agents with 80% of the agents initially using the strategy with delay $\tau = 5$. Plots show the fraction $f$ of agents using resource 1 as a function of time. To insure that each strategy always maintains a minimal population, we modify the reward mechanism so that agents cannot switch out of strategies that have less than five agents. Agents continually switch strategies in trying to achieve local improvements. As a result, the system sometimes goes unstable and the equilibrium is punctuated by bursts of activity.

As explained above and illustrated in figure 1, the strategy switching mechanism leads to a stabilizing strategy mix. This strategy mix, however, is not unique. Rather, there is a repertoire that is consistent with a stable equilibrium. Thus once the system equilibrates agents continually switch strategies in order to optimize their own performance. This is achieved without destroying the overall system stability with respect to resource utilization. This implies an effective degeneracy whereby a number of strategy mixes are consistent with the observed equilibrium. At the dynamical level, this degeneracy manifests itself in an effective random walk in the space of strategies. As time goes on this random walk can occasionally push the strategy mix into values that are not consistent with stable resource contention. This results in the observed bursts of activity, which are subsequently quenched by the Hogg-Huberman mechanism.

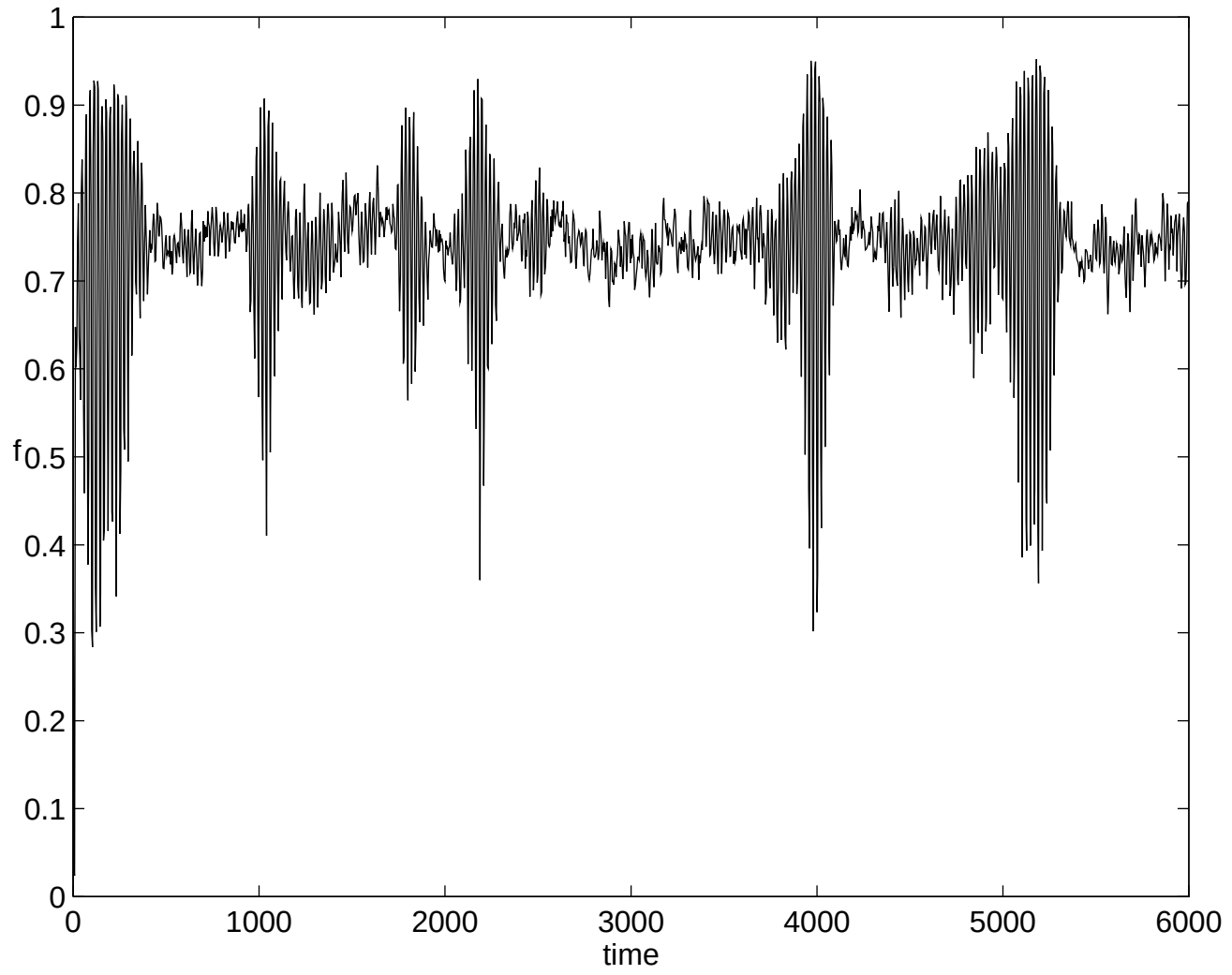

**Fig. 4.** Typical behavior of a system with 5000 agents when particular strategies involve preferences towards specific resources. Resource utilization information is delayed by $\tau = 10$ and the payoffs and parameters are as in figure 1. In contrast to the case of purely time delayed strategies, where different strategy mixes either stabilized or unstabilized the same resource utilizations, different mixes of preferences correspond to different equilibria in resource utilization.

In order to study this phenomenon in greater detail, we consider the simpler case in which the agents are limited to three strategies. Since we need to guarantee that there are indeed strategy mixes that stabilize the resource contention, we limit one of the strategies to have zero delay. We choose the other strategies to have delays $\tau = 3$, and $\tau = 5$, and set the payoffs, uncertainty $\sigma$, and the rate $\alpha$ in the system to be as before.

Figure 5 shows the results of a simulation with 2000 agents for two values of the strategy reevaluation rates, $\gamma = 0.01$ and $\gamma = 0.2$. The observed behavior clearly shows a stable equilibrium punctuated by bursts of instability. Note that for the reevaluation rate $\gamma = 0.01$ bursts occur less frequently than in the case of the higher value of $\gamma$, but when they do occur they tend to relax back in slower fashion than in the case of high reevaluation rate. There is thus a trade-off in the rate of the reevaluation rate: for higher $\gamma$, the system relaxes back to equilibrium faster, but also tends to go unstable more frequently; for smaller $\gamma$, the system will take a longer time to relax back to equilibrium, but also tends to stay stable longer.

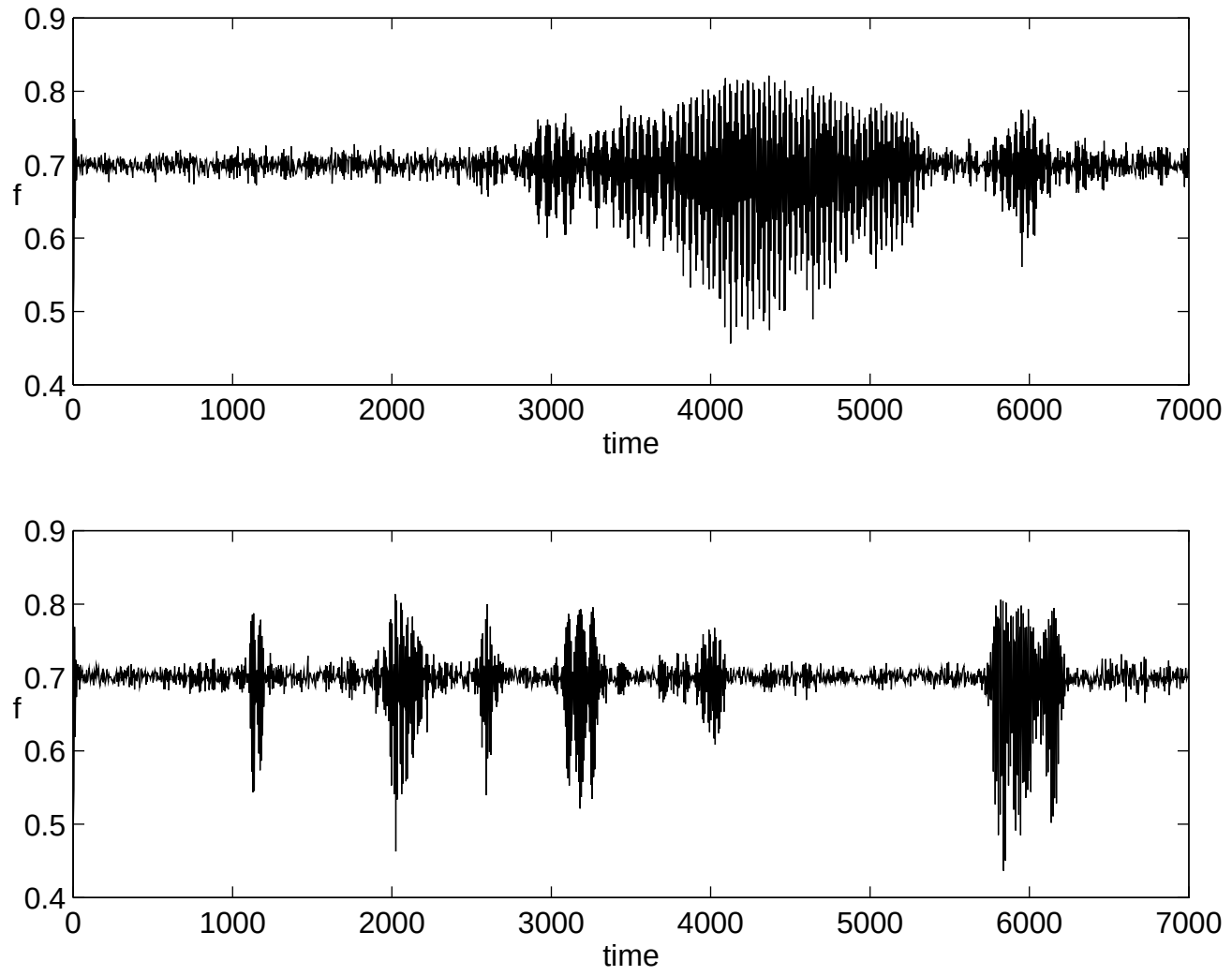

**Fig. 5.** Computer simulations of contention over two resources with strategy switching among 3 strategies. Simulations consisted of 2000 agents total. System parameters and payoffs are as in figure 1. The three strategies have associated delays of 0, 3, and 5 time units respectively. Top figure shows results from using a switching rate $\gamma = 0.01$. Note again the stability punctuated by bursts of activity. Bottom figure shows results for a faster switching rate $\gamma = 0.2$. In this case bursts occur more frequently but have shorter durations when they occur.

# 4 A Theory of Bursts

### Stability Properties

In this section we discuss the stability of resource contention for different strategy mixes. The results of the previous section show that performance driven strategy switching provides a mechanism through which the distribution of strategies stabilizes the resource contention in the system. But the mix of strategies cannot be totally arbitrary, for there will be strategy mixes for which the system will become unstable. In order to determine the range of strategy mixes that are consistent with stability, one needs to perform a linear stability analysis around the equilibrium point generated by the dynamics. We here outline the results of such an analysis [17]. Small perturbations away from equilibrium evolve in time as $e^{\lambda t}$. This means that for positive, real values of $\lambda$ the perturbation will grow in time, and for $\lambda$ negative it will decay, relaxing back to equilibrium.

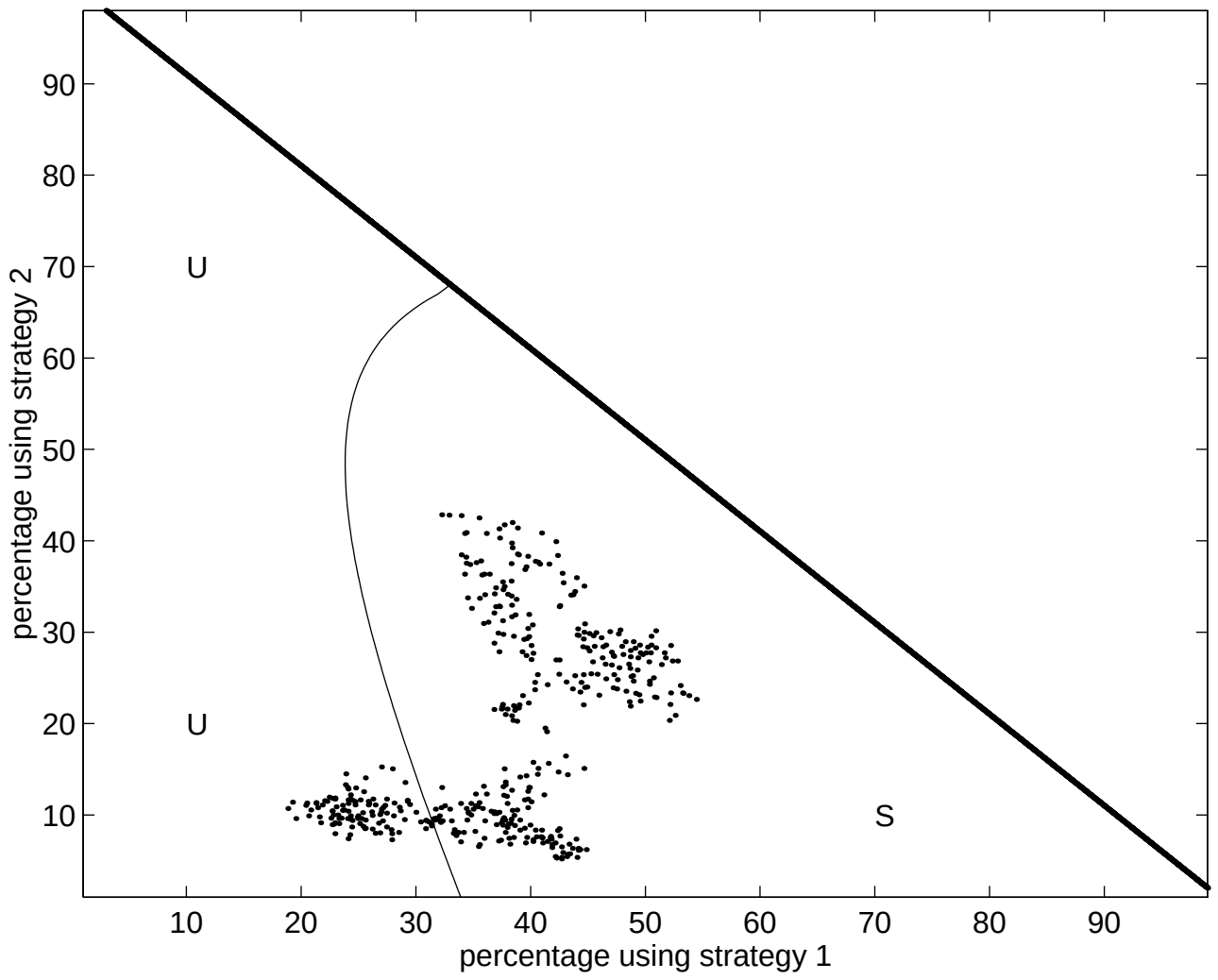

**Fig. 6.** Stability of regions in strategy space for parameters and delays used in figure 5. The horizontal axis denotes the percentage of agents using zero delay, $\tau = 0$ and the vertical axis denotes the percentage of agents using delay $\tau = 3$ . Regions of strategy space that stabilize resource usage are denoted by S and unstable regions are denoted by U. Also shown are the points in strategy space mapped out by the $\gamma = 0.2$ simulation in figure 5 from before a burst occurs until after its occurrence (from time units 4800 to 6600). Strategies randomly move around the stable region while resource contention is still stable. Once the strategies cross into the unstable region, a burst of activity occurs, and the strategy mix is pushed back into the stable region.

For a given strategy mix, $f_s^{str}$, $\lambda$ is determined by the following equation [17],

$$\lambda + \alpha - \alpha \sum_s f_s^{str} \rho_s'(f_{eq})\, e^{-\lambda \tau_s} = 0 \tag{9}$$

Here $\rho_s'$ is the derivative of the function, $\rho_s$ which is given by eqn. 4 evaluated at the relevant fixed point, $f_{eq}$. This fixed point is determined by

$$f_{eq} = \sum_s f_s \rho_s(f_{eq}). \tag{10}$$

In eqn. 9, $\tau_s$ is the delay relevant to strategy *s*, and $b_s$ is the preference towards the use of a given resource by strategy *s*. A particular strategy mix will therefore lead to stable resource allocation if all the solutions (in general there are an infinite number of them) to eqn. 9 have negative real parts. This will guarantee that fluctuations away from the fixed point in resource contention will eventually relax back to the equilibrium. Notice that when strategies are purely unbiased and thus specified only by a delay, the fixed points in resource utilization are the same as those in which information is not delayed.

Since each strategy mix will either lead to stable or unstable resource contention, the strategy space can itself can be broken up into two regions: a region containing strategy mixes for which the resource allocations are stable, and another containing strategy mixes for which resource allocations are unstable.

Figure 6 shows the regions of *strategy space* that support stable and unstable resource contention for the parameters and strategies as in figure 5 (the value of $\gamma$ does not affect these regions). In this case, since there are 3 strategies ($\tau = 0,\ 3, \text{and}\ 5$), and since the fractions of agents using all strategies must add up to one, each strategy mix is uniquely defined by the fraction of agents using the first two strategies with delays $\tau_1 = 0$ and $\tau_2 = 3$. Recall that a strategy mix is in the stability region if solutions for $\lambda$ in eqn. 9 exist that have negative real part. Note the rather large region in the strategy space for which resource allocation is stable. Superimposed on this figure is the evolution of the strategy mix for the simulation shown in the lower part of figure 5.

Note that when the strategy mix is in the stable region, it seems to behave randomly with no particular bias. Agents switch strategies with no penalty since all strategies perform equally well. As a result, small fluctuations (in this case induced by having a finite number of agents) make the strategy population execute a random walk within the region of stability. However, after the strategy mix moves into the unstable region, resource contention becomes unstable. Once the resource contention is destabilized, strategy switching is once again biased towards enhancing strategies that perform well. The strategy switching mechanism then forces the mix back into the region of stability, thus stabilizing the resource contention in the system. This process then accounts for the characteristic bursts found in the computer experiments that we described above.

The fact that strategies tend to perform equally well when the overall system is in equilibrium is related to the fact that the strategy switching mechanism in the system is highly adaptive. This adaptability not only allows the system to automatically search out strategy mixes that stabilize resource utilization, but also manifests itself in a degeneracy of solutions for which resource utilizations are stable. We discuss this degeneracy in the next section.

## 5 Degeneracies in Multiagent Systems

Since our model is intended to capture general features of an adaptive agent system, we assume that the agent population is not "preprogrammed" towards any particular distribution. We call this assumption *neutral adaptability*. This is the case for the mechanism given by eqn. 5. Rather, the agent population adapts with respect to the external environment and variations in the resource payoffs, leading to strategy mixes that stabilize the system. There are, however, many strategy mixes that stabilize the resource contention. It is precisely this degeneracy in the stabilizing mixes that makes the system susceptible to bursts of volatility characterized above.

In order to better examine this degeneracy, we reparametrize the system equations for $f_{rs}$ ($r = 1, ..., N_r,\ s = 1, ..., N_s$) so that , $\eta_s$ couples directly to only $N_s$ equations rather than to the $N_s N_r$ such equations as given by eqn. 6. This reparametrization is given by $f_r^{res}$, $f_s^{str}$, and

$$F_{rs} = f_r^{res} f_s^{str} - f_{rs}, \tag{11}$$

subject to the consistency conditions,

$$\begin{cases} \sum_r & f_r^{res} = 1 \\ \sum_s & f_s^{str} = 1 \\ \sum_s & F_{rs} = \sum_r F_{rs} = 0 \end{cases} \tag{12}$$

With these conditions the number of degrees of freedom for the corresponding parameters are given by:

$$\begin{cases} f_r^{res} : & (N_r - 1) \\ f_s^{str} : & (N_s - 1) \\ F_{rs} : & (N_s - 1)(N_r - 1) \end{cases} \tag{13}$$

Note the system has a total of $N_r N_s - 1$ degrees of freedom. Together, these quantities completely specify the system space for the strategy-resource system, leading to the following dynamical equations,

$$\begin{cases} \frac{dF_{rs}}{dt} = -(\alpha+\gamma)F_{rs} + \alpha f_s^{str}\left(\sum_{s'} f_{s'}^{str}\rho_{rs'} - \rho_{rs}\right) \\ \frac{df_r^{res}}{dt} = -\alpha\left(f_r^{res} - \sum_s f_s^{str}\rho_{rs}\right) \\ \frac{df_s^{str}}{dt} = -\gamma\left(f_s^{str} - \eta_s\right) \end{cases} \tag{14}$$

With these expressions for the dynamics, we now turn to a study of the degeneracy in the system that is useful in analyzing the occurrence of bursts.

## Neutral Adaptability

We now make explicit a notion of *neutral adaptability*. First, we assume the existence of a nonempty set, $E_r$, consistent with the equilibration of the resource utilization $f_r$ ($\frac{df_r^{res}}{dt} = 0$) and the equilibration of $F_{rs}$ ($\frac{dF_{rs}}{dt} = 0$). In terms of the second equation from the triplet above this implies that,

$$f_r^{res} - \sum_s f_s^{str}\rho_{rs} = 0 \tag{15}$$

We now define a system to be *neutrally adaptable* with respect to the set $E_r$ if it can settle into strategy mix equilibria contained in $E_r$. From the right hand side of the last equation in the triplet above we can state this mathematically as $\eta_s = f_s$ on $E_r$. Since this means that

$$\frac{df_s^{str}}{dt} = 0 \tag{16}$$

on that set, the agent population is macroscopically unbiased towards any particular population of strategies on such a set. In this way, if there exists a strategy distribution that stabilizes resource contention, the system can automatically adapt to that distribution and stabilize the entire system without any “preprogramming”.

**Delays and Preferences** We now discuss the specific form for $\eta_s$ given by eqn. 5 for the case of time-delayed strategies. In this case, the relevant set $E_r$ is given by the constraints

$$\begin{cases} f_r^{res} = \rho(f_r^{res}) \\ F_{rs} = 0 \end{cases} \tag{17}$$

This set forms an $N_s - 1$ dimensional manifold that can be parametrized by the values of $f_s^{str}$, in which the probability that strategy *s* is chosen by an agent is

$$\eta_s = \frac{\sum_r \left(f_r^{res} f_s^{str} - F_{rs}\right) G_r}{\sum_r f_r^{res} G_r} = f_s. \tag{18}$$

More intuitively, neutral adaptability is here a result of the fact that strategy switching does not differentiate between different time delayed strategies when the contention over resources with degrees of freedom specified by $f_r^{res}$ and $F_{rs}$ have relaxed to an equilibrium.

When strategies consist of preferences towards different resources, the relevant set $E_r$ is determined instead by the constraints,

$$\begin{cases} G_r(f_r^{res}) = G_{r'}(f_{r'}^{res}) \quad \forall r' \\ f_r^{res} = \sum_s f_s^{str} \rho\left(\frac{b_s}{\sigma}\right) \\ \frac{dF_{rs}}{dt} = 0 \end{cases} \tag{19}$$

This is a $N_s - 2$ dimensional manifold that once again satisfies the neutrality condition,

$$\eta_s = \frac{\sum_r \left(f_r^{res} f_s^{str} - F_{rs}\right) G_r}{\sum_r f_r^{res} G_r} = f_s^{str} - \sum_r F_{rs} = f_s^{str}. \tag{20}$$

Strategy switching among different preferences does not differentiate between strategy distributions that hold the system at resource allocations that have equal payoff.

Since neutrally adaptive switching mechanisms do not differentiate between different strategy mixes, it follows that sections in the system space are degenerate and that surrounding regions are approximately so. In these regions the strategy population continuously changes its composition and approaches stability boundaries without any corrective mechanism. In turn, the stability of such systems does not increase as rapidly with agent population size as might be expected from standard fluctuation corrections [21] to the meanfield approximation eqn. 6.

**Discussion** Having established how neutral adaptability is associated with clustered volatility we now point out that it is only a sufficient, but not necessary condition. For example, consider the case in which agents switch between delayed strategies randomly with probability $\beta$ and according to the switching mechanism $\eta_s$ with probability $1-\beta$. In this case the last equation in the triplet eqn. 14 is changed to

$$\frac{df_s}{dt} = -\gamma\left(f_s - (1-\beta)\frac{\sum_r f_{rs}^{res} G_r}{\sum_r f_r^{res} G_r} - \frac{\beta}{N_s}\right) \tag{21}$$

Similarly to the previous analysis, we consider the set $E_r$ given by 17 that consists of equilibrium configurations for the degrees of freedom $f_s$ and $F_{rs}$. On this set, the second term in the right hand side of 21 reduces to $(1-\beta)f_s$as in the neutrally adaptive case, and becomes

$$\frac{df_s}{dt} \underset{E_r}{=} -\gamma\beta\left(f_s - \frac{1}{N_s}\right) \tag{22}$$

This equation shows the existence of an entropic 'force' that pushes the strategy mix towards a population of equally populated strategies, i.e. $f_s = \frac{1}{N_s}$. Assuming that $\beta \ll 1$ this force will be prominent only when the system is within the stability boundary. If we assume that the uniform strategy population lies outside the stability boundary for resource contention, then this entropic contribution will tend to enhance the generation of bursts, and hence volatility clustering.

Therefore, *neutral adaptability* is not a necessary condition for the appearance of clustered volatility. Rather, *neutral adaptability* is a convenient assumption for studying volatility clustering. Without this assumption, resource payoff functions would require a certain amount of tuning for the phenomenon to appear. It is the degeneracy in strategy mixes that stabilize resource contention that plays a central role. This degeneracy is a result of the fact that boundedly rational agents cannot discern between strategies when the system is close to stability. As a result the strategy switching mechanism that dramatically stabilized the system becomes less effective resulting in clustered bursts of activity.

## 6 Conclusion

In this paper we have studied the dynamics of a multiagent system whose members continuously modify their behavior in response to the dynamics that unfold within the system. Their behavior is driven by local optimization of the utilities that agents accrue when gaining access to resources. Agents decide on the basis of having bounded rationality, which implies imperfect information and delayed knowledge about the present state of the system.

Given these constraints we have shown that with the introduction of a set of diverse strategies, an initially unstable system can flow towards equilibrium. But within this equilibrium, fluctuations among possible strategy mixes that are consistent with the fixed point lead to the phenomenon of volatility clustering. Since many possible distributions of strategies are consistent with equilibrium, as time progresses the agent population explores different strategy mixes. These underlying fluctuations have the structure of a random walk in the space of strategies and show up even when there are large number of agents. If a strategy mix moves beyond the boundary of stability there results bursts of momentary instability. These bursts eventually relax into a stable equilibrium. This mechanism of stability punctuated by clustered volatility appears to be quite general in systems where agents explore strategies in search of local improvements.

In order to analytically understand the appearance of these bursts, we established the stability boundaries and the nature of the degeneracies in the system. We also introduced the notion of *neutrally adaptive* strategy switching which provides a sufficient condition for the appearance of volatility clustering. We also discussed the existence of other mechanisms that can lead to the same phenom-enon, such as the existence of an entropic mechanism that leads to a system evolution whose asymptotic behavior is characterized by a uniform sampling of all the strategy mixes of the system.

We also suggested that this new phenomenon may explain the clustered volatility observed in some economic time series, such as stock returns and exchange rates. In the case of high inflation, it has been pointed out that the daily fluctuations of currency exchanges exhibit the same bursty behavior that we discovered [22]. It is not clear however, that our mechanism works in that context, for it has been pointed out that in inflationary times such bursts might be the result of amplified system responses to exogenous fluctuations.

This work also offers a complementary view of clustered volatility to that recently presented via computer simulations of adaptive agents by Grannan and Swindle [23]. The formalism we have presented may also be helpful in explaining their results on a more analytical basis. In addition, the explicit introduction of dynamical equations allows for a determination of the nature of the equilibrium, as well as the parameter values for which this phenomenon is likely to appear. Moreover, it opens the way to study more complicated sets of strategies.

Finally, there remains the interesting problem of resource contention in computer networks, where the introduction of servers with increased and specialized capabilities will lead to a desire on the part of users to access them, thereby creating long queues and increased costs. If computational agents are programmed to take advantage of local fluctuations in the use of these resources, the results of this paper might have implications in the design and implementation of widely distributed computational systems.

## Acknowledgments

We have profited from discussions with Tad Hogg and Scott Clearwater. This work was partially supported by the Office of Naval Research under contract No. N000014–82–0699.